# Blocks-in-Conduit: REBCO Cable for a 20T@20K Toroid for Compact Fusion Tokamaks

P.M. McIntyre *IEEE,* J. Rogers, and A. Sattarov

*Abstract*— Blocks-in-Conduit is a novel approach to cable, coil, and splice technologies with unique benefits for a high-current-density toroid winding to operate at 20 T at 20 K. Blocks of REBCO tape are cabled in conduit, with a laminated structure and thin-wall center-tube that provides twist along the cable, spring-loaded stress management within the cable, and cross-flow cooling throughout a thick toroid winding. The coil technology utilizes a co-wound armor structure that integrates stress management, cross-flow cooling and bypass of coil stress to protect the BIC. An interleaved splice joint enables low-resistance demountable splices within the winding pack. These provisions yield maximum current density in the winding pack, maximum stability of the REBCO tape blocks, and minimum conductor cost for a tokamak toroid.

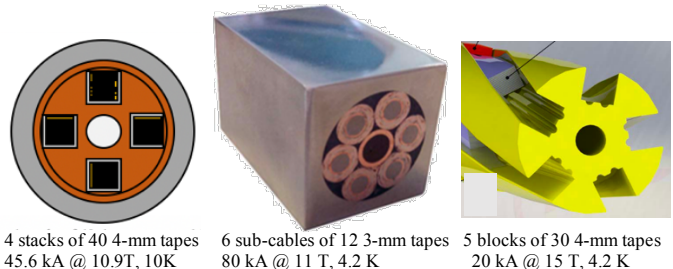

Fig. 1. REBCO cables for fusion tokamaks: a) MIT/CFS's VIPER; b) ACT's CORC; c) ENEA's twisted-stack cable.

- 4 stacks of 40 4-mm tapes — 45.6 kA @ 10.9T, 10K
- 6 sub-cables of 12 3-mm tapes — 80 kA @ 11 T, 4.2 K
- 5 blocks of 30 4-mm tapes — 20 kA @ 15 T, 4.2 K

## I. Introduction

**M**ENARD [1] examined the impact of magnet systems and core geometry for magnetic confinement of a plasma capable of net-positive fusion power production. He identified the strategic importance of increasing the field strength of the confining magnetic field to $B_0 > 16$ K, increasing the overall current density in the windings to $J_W > 80$ A/mm$^2$, and operating the superconducting windings at $T_0 > 20$ K. As context, the same parameters for the toroid of ITER are 12 T, 17 A/mm$^2$, and 5 K respectively. The winding current density >80 A/mm$^2$ required for effective fusion poses a major challenge to the present state-of-art for superconducting wire, cables, and windings.

REBCO is the only superconductor that can provide high current density in high field at temperatures in the range 10-40 K, and so it is the conductor of choice for the solenoids and toroids required for magnetic-confinement fusion. But high-field windings of large size require an armored conductor capable of >40 kA operating current. REBCO can only be fabricated in thin tape of 1 cm width with tape current ~300 A at high magnetic field. Bruzzoni *et al*. [2] summarize the several requirements for a REBCO-based cable to be used in the windings of fusion magnets:

- the cable must incorporate ~200 tapes;
- tapes must be supported in a matrix that can provide low-resistance current-sharing as the cable current is ramped;
- tape stacks must be twisted along the cable so that all tapes carry the same current everywhere in the winding;
- the cable must be reliably insulated and supported against crushing Lorentz stress.

Those provisions are not easy to achieve. The MIT/CFS VIPER cable [3] (Fig. 1a), ACT's CORC cable [4] (Fig. 1b), and ENEA's twisted-stack CIC [5] (Fig. 1c) each provide twist pitch to sustain current distribution, but there are challenges to achieve uniform face-face compression of tapes within the CORC cable, and in the VIPER cable all tapes are soldered and so pose challenges for bending to form toroid windings.

A collaboration of Accelerator Technology Corp. and Texas A&M University is developing a Blocks-in-Conduit (BIC) conductor, shown in Fig. 1c, that integrates stress management, current-sharing, and cross-flow cooling throughout the cable itself. The cable is designed to operate with ~40 kA in 20 T field at 20 K temperature. This paper presents the cable design, specifics of its fabrication, and calculations of its expected performance for the above considerations.

## II. BIC fabrication

Fig. 2 shows the sequence of steps in fabricating a Blocks-in-Conduit (BIC) cable [6]:

First thin copper laminations (Fig. 2a) are prepared by cutting the desired shape from copper sheet using either wire-EDM or die-stamping. Each lamination contains 4 rectangular channels that hold the tape blocks, a center hole, and 4 alignment slots.

The laminations are stacked on a rail fixture, and aligned using a set of rods in four slots to precisely align the lamination stack (Fig. 2b). A perforated stainless-steel center tube is inserted through the center hole of the lamination stack.

P.M. McIntyre is with the Accelerator Research Lab, Texas A&M University, College Station, TX 77845 USA and also with Accelerator Technology Corp. (e-mail: p-mcintyre@tamu.edu).
J.S. Rogers is with Accelerator Research Lab, Texas A&M University, College Station, TX 77845 USA (email: jsr12e@tamu.edu).
A. Sattarov is with Commonweath Fusion Systems, Boston, MA USA (e-mail: akhdiyorsattarov@gmail.com).

Color versions of one or more of the figures in this paper are available online at http://ieeexplore.ieee.org.
Digital Object Identifier will be inserted here upon acceptance.

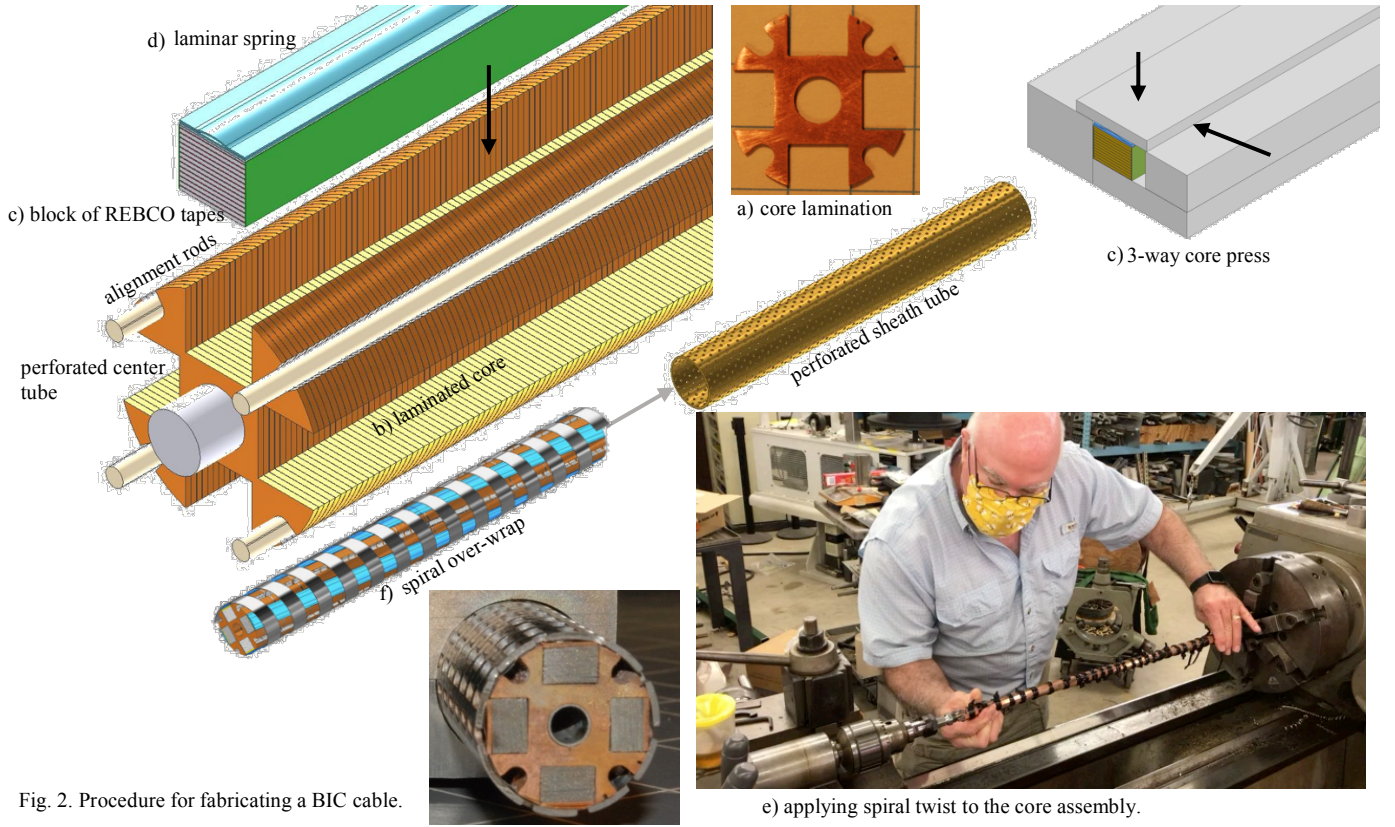

Fig. 2. Procedure for fabricating a BIC cable.

e) applying spiral twist to the core assembly.

A stack of ~50 REBCO tapes is assembled on a 3-way stacking press (Fig. 2c) that aligns one edge of all tapes to one side of the press. Two thin strips of fluxed low-melt solder foil are located facing the side edges of the tape block; a laminar spring (Fig. 2d) is located on the top face of the tape block and compressed to preload the tape block; and the solder strips are pressed into the side faces of the tape block to secure them as a subassembly.

Tape block subassemblies are inserted as a close fit into the four rectangular channels in the lamination stack (Fig. 2c), and a set of tie-wraps is secured on the cable surface to temporarily hold the package intact. The dimensions are chosen so that once the cable is compacted the spring will provide a ~1 MPa linear compression of the tape stack.

The cable is mounted on a rotary mandrel and twisted to create a desired twist pitch (Fig. 2e). The twist pitch interchanges each tape-stack within the cable from inside to outside in the winding, and its value $\lambda$ can be chosen to cancel internal strain that would otherwise be created when the cable is bent on a radius of curvature R: $\lambda = N\pi R/2$ for some integer N. The laminated structure is quite flexible, and the cable can accommodate bends with R~60 cm.

The tie-wraps are removed after the cable is twisted, the core assembly is tension-wrapped with a spiral over-wrap of thin stainless-steel, and the ends of the over-wrap are spot-welded to the lamination stack to lock in the pre-load (Fig. 2f). The core assembly is inserted as a loose fit into a perforated sheath tube and the sheath tube is drawn down onto the core assembly to compress the laminar springs and lock in the twist pitch. The BIC cable is then complete and ready to co-wind for a magnet winding.

## III. CURRENT-SHARING IN BIC CABLE

Current-sharing is critical in a REBCO cable. When a tape is transposed within the cable the orientation angle θ between the tape face normal and the direction of the magnetic field at its location in a winding changes from parallel (θ =90°) where $I_\parallel$ is maximum and perpendicular ((θ =0°) where $I_\perp$ is minimum. The ratio is large ($I_{c90}/I_{c0}$>3 for 20 T, 20 K) and so current must be able to transfer with low normal-state resistance from tapes of one block to tapes of another block everywhere along the cable. This requirement is difficult to achieve in a cable that has the flexibility to bend with significant curvature.

The REBCO tapes are manufactured with a copper cladding on all surfaces, so that the contact among tapes in a tape block is made through the contact resistance $R_c$. Lu et al. [7] measured the contact resistance between faces of Cu-clad REBCO tapes as a function of the compressive stress under which tapes are compressed. If the interface among tapes is sustained with compression $S_c$ > 1 MPa, then the contact resistance between the copper cladding of REBCO tapes is $R_c \sim 35~\mu\Omega \cdot cm^2$.

Each tape block in a BIC cable is compressed by a laminar spring to sustain this critical value of $S_c$, when the cable is fabricated, when it is co-wound on curvature in a winding, and when it is in cryogenic operation at high magnetic field.

Current must also transfer with low resistance from block to block within the cable. Provision is made for this purpose to produce a fluxed solder bond between the edges of the tapes and the side walls of the rectangular channel in the lamination stack. Strips of fluxed low-melt solder are pressed into the edges of the tapes during assembly of the cable. After the cable is co-wound into its final form in a winding, a current is pulsed through the cable at room temperature, sufficient to heat the core assembly to the flow temperature of the solder. The tapes are thereby bonded in their rectangular channels in the lamination stack and thereby provide reliable low-resistance normal-state current-sharing among the four conductor blocks throughout the winding.

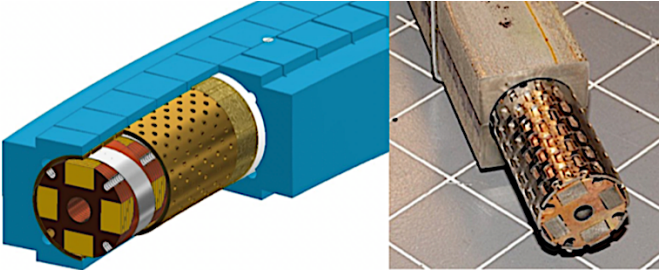

Fig. 3. Co-wound armor as it is assembled on a segment of BIC cable.

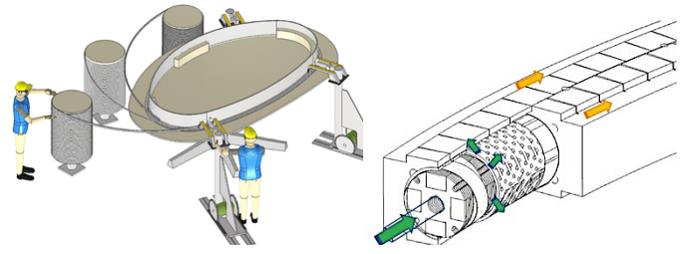

Fig. 4. a) Co-winding of BIC cable and armor half-shells onto a toroid winding; b) cross-flow cryogen cooling within the BIC cable and co-wound armor.

## IV. CO-WOUND ARMOR

A toroid or solenoid winding for a fusion magnet must support the cables in the presence of an accumulation of Lorentz stress through a thick winding package. Conventionally a high-strength armor sheath is compressed onto the cable as it is manufactured, then the armored cable is bent to the curvature required for a winding. In such assemblies the compression of armor onto the cable typically damages some segments of wires and leaves other segments un-supported [8].

BIC cable is supported by co-winding two continuous half-shells of high-strength alloy armor so that they conform to the BIC cable with the desired curvature radius R of the winding. The assembly of armor half-shells provides robust support of both hoop stress and transverse stress within the winding, and the armor half-shells bypass the accumulation of Lorentz stress produced by other layers of the winding so that they do not produce strain inside the BIC cables.

Fig. 3 shows a segment of co-wound armor, in which each armor half-shell is fabricated as a straight channel of high-strength metal alloy, for example Haynes 620, with a half-cylindrical inner contour, two rectangular step channels, and a linear array of kerf-cuts. The cut width and the spacing of the kerf-cuts are chosen so that, when the BIC cable and armor half-shells are co-wound onto a section of a toroid winding with curvature radius R, the inner armor clamshell bends inward at the bottom of the kerf-cuts so that the outer edge of adjacent kerf-cuts nearly closes. The kerf-cuts thereby relieve the bending strain that would otherwise be created if one were to bend a thick structural beam, so that each armor half-shell retains its full structural strength to support the bridging of radial stress within a multi-layer winding, and also the web portion of each half-shell (which is not kerf-cut) retains its full strength to support hoop stress in each layer.

Fig. 4a shows a configuration for co-winding the BIC cable and its armor shells under tension onto a toroid winding. Strips of mica paper are inserted in the annular gap between the outer surfaces of the perforated sheath tube and the inner clamshell surfaces of the twos armor half-shells. The mica paper provides a low-friction slip-surface between the BIC cable and the inner and outer cylindrical surfaces of the armor half-shells.

As successive turns of armored cable are wound, the outer armor half-shells of adjacent turns are spot-welded together to secure the conformation of each turn. Interlayer insulation is provided by applying a laminar assembly of mica paper, fiber-ceramic fabric, and a slurry of low-melt glass frit.

## V. CROSS-FLOW COOLING

In a magnetic-confinement fusion device that drives D-T fusion to make net electric power, the windings must operate in an environment of intense heat and fast neutron fluence. Even with an optimized cryostat, it will be important to provide cryogenic heat transfer throughout the entire winding, so that heat that is delivered into the interior of a thick winding can be removed with minimum temperature differential.

Fig. 4b shows a configuration of the co-wound armor that provides for cross-flow of coolant fluid throughout the entire length of every turn of BIC cable in a winding. Coolant flow is sustained by providing two manifolds at one location on the circumference of the winding. A supply manifold injects the supply flow into the center tube of all turns of the BIC cable; and a return manifold returns the return flow from the rectangular step channels at the four corners of the armor half-shells. Radial channels are provided in periodically spaced laminations to provide a radial fluid flow between the perforated center tube and the rectangular channels. The flow pattern is adjusted to provide a uniform volumetric heat transfer along the entire length of all turns in the winding.

## VI. BARREL WINDING AND GRADED CABLE COMPOSITION

The maximum superconducting current $I_c(B)$ that can be carried by an HTS tape depends strongly upon the magnetic field strength B that is ambient in its location in the winding. Fig. 5a shows a BIC-based toroid for a compact spherical tokamak [9], designed to operate at 20 T maximum field, 20 K temperature. Fig. 5b shows a cross-section of one segment winding, which is wound in four 2-layer sub-windings, connected in series with cable current $I_0$. The magnetic field $B_n$ in succeeding sub-layers decreases with $n$, (Fig. 6a) and the number of tapes in the BIC cable for each sublayer is chosen so that $I_0 = N_n I_c(B_n)$. The table in Fig. 5 summarizes the parameters for each sub-winding. Barrel-winding with graded cable composition reduces by half the quantity of REBCO tapes required for a given toroid design.

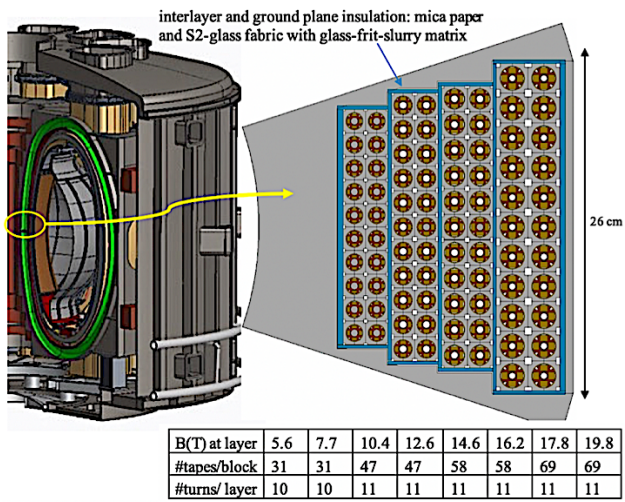

| B(T) at layer | 5.6 | 7.7 | 10.4 | 12.6 | 14.6 | 16.2 | 17.8 | 19.8 |
|---|---|---|---|---|---|---|---|---|
| #tapes/block | 31 | 31 | 47 | 47 | 58 | 58 | 69 | 69 |
| #turns/ layer | 10 | 10 | 11 | 11 | 11 | 11 | 11 | 11 |

Fig. 5. a) Toroid winding for a 20 T compact spherical tokamak; b) cross-section of the winding showing 4 sub-windings with graded composition.

## VII. Stress Management, winding current density

The BIC cable presented above provides stress management at the cable level, and the co-wound armor provides stress management throughout the winding. Fig. 6b shows the calculated distribution of von Mises stress in the BIC cables and in the co-wound armor. Fig. 6c shows the calculated von Mises stress in the superstructure that supports the configuration of 10 toroid segments. The benefit of the co-wound armor in bypassing the stress in the overall winding so that it does not compromise the tape blocks within each BIC cable is evident in both distributions.

## VIII. Interleaved splice technology

The internal stress management in a BIC cable makes it possible to make interleaved splices joints in which the tapes of each tape block are interleaved within the spiral channel. The splice design is summarized in Fig. 7. The sheath and washer stack at the two ends of a cable segment are cut back from the ends of the tape blocks, a special washer stack is installed to connect the ends of the cable structures (Fig. 7a) and mechanically linked into the ends of both cables (Fig. 7b). The end segments of all tapes are solder-tinned over a 10-cm end, fanned out radially, and then interleaved into the deeper channels of the splice washer stack. A helical laminar spring is installed on the outer face of the interleaved stack, and a spiral over-wrap is applied to compress the laminar springs, and the joint is heated to flow the fluxed solder and bond all overlapping tapes within each of the block.

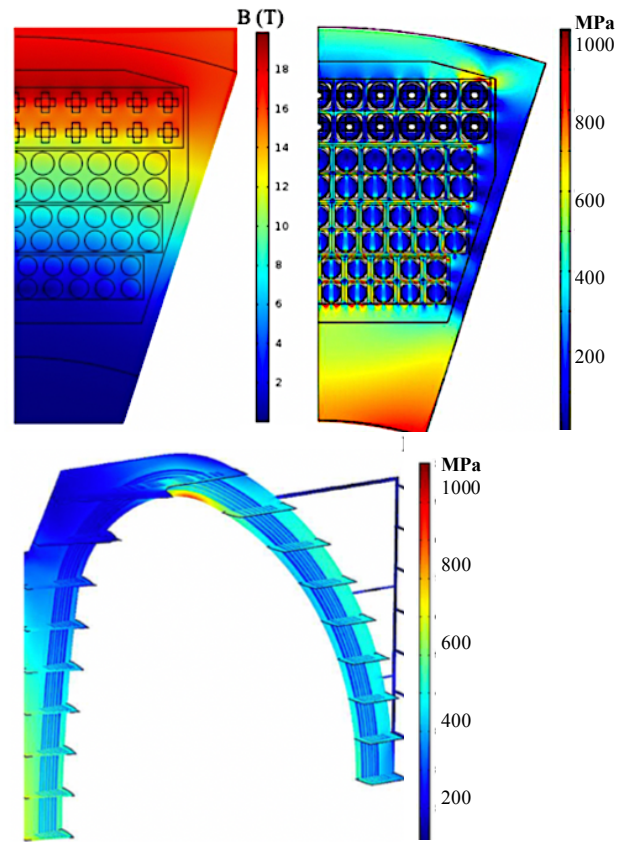

Fig. 6. a) Magnetic field distribution in the winding; b) von-Mises stress distribution in the BIC cables and co-wound armor; c) von Mises stress distribution in the support superstructure.

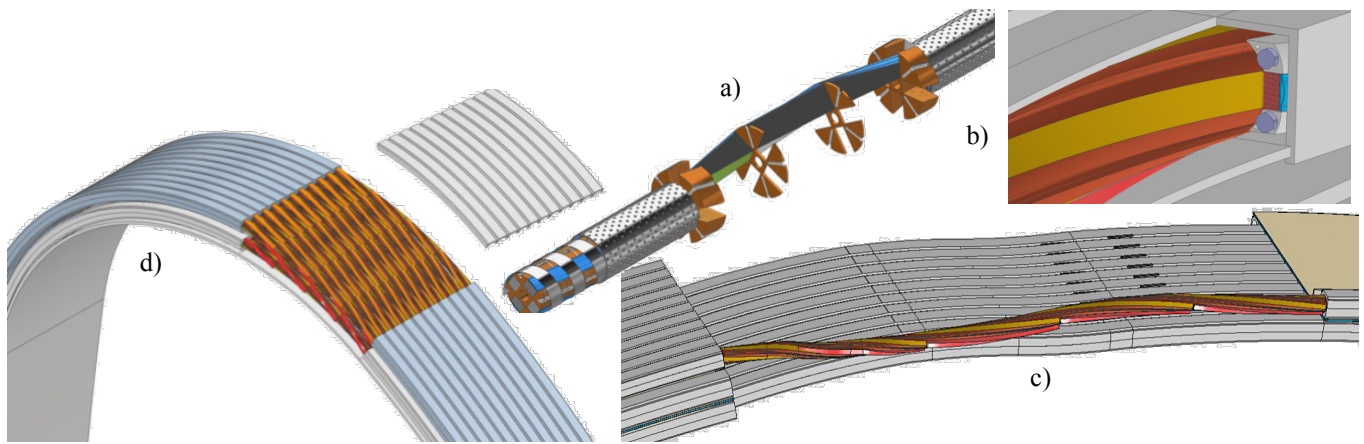

Fig. 7. Interleaved splice technology for BIC cable: a) assembly of 50 interleaved tapes of one block, supported in the rectangular channel of the splice core; b) integration of the splice core with the armor of the two cables to provide stress bridge in the joint region; c) serial string of 4 splice joints space l/4 pitch apart so they are all accessed from one direction; d) splicing of all turns in a winding can accommodate installation/removal of the winding on the plasma vessel of a toroid.

The four blocks within a cable may be spliced at locations that are space a distance λ/4 apart so that all 4 splices are made on one face of the armored cable. This provision makes it possible to prepare and bond all splices from one accessible orientation to that cable turn, as shown in Fig. 7c.

The motivation of this approach is to develop a robust splice joint with such low splice resistance that it could be used to splice all turns of each winding of a toroid (Fig. 7d). It would then be possible to assemble each winding onto the plasma vessel and later replace it without disassembling the plasma vessel, similar to the jointed windings of NSTX-U.

### Conclusions and Plans for Future Development

The BIC cable is designed to provide optimum twisting and current-sharing within a high-current-density REBCO cable. The co-wound armor should provide stress management at both cable level and winding level, and facilitate layer-winding of a high-field toroid. With layer-winding it should be possible to reduce by half the quantity of expensive REBCO tape, and to achieve an overall winding current density of ~100 A/mm$^2$, meeting the requirement for net fusion power in the analysis of Menard [1].

We have successfully fabricated a 60 cm long segment of BIC cable (Fig. 2e, Fig. 3b) and validated the fabrication methods. We plan to fabricate a 3 m segment with interleaved splices on both ends, bend it into a hair-pin specimen. We plan to then test the specimen in a test facility such as SULTAN.

### Acknowledgments

The authors would like to thank Thomas Brown, Steven Cowley, and Michael Sumption for many helpful discussions that helped us to develop an understanding of the interplay of requirements for the windings of tokamaks.

### References

[1] J.E. Menard, 'Compact steady-state tokamak performance dependence on magnet and core physics limits', Phil. Trans. R. Soc. **A377** (2019) 20170440.
[2] P. Bruzzone *et al.*, 'High temperature superconductors for fusion magnets', Nucl. Fusion **58** (2018) 103001.
[3] Z.S. Hartwig *et al.*, 'VIPER: an industrially scalable high-current high-temperature superconductor cable', Supercond. Sci. Technol. **33** (2020) 11LT01.
[4] D.C. van der Laan, J.D. Weiss, and D.M. McRae, 'Status of CORC cables and wires for use in high-field magnets and power systems a decade after their introduction', Supercond. Sci. Technol. **32** (2019) 033001.
[5] G. Celentano *et al.*, 'Design of an industrially feasible twisted-stack HTS Cable-in-Conduit conductor for fusion application', IEEE Trans. Appl. Superconduct. **4**, 3, 4601805 (2014).
[6] The Blocks-in-Conduit technology is the subject of a provisional patent application, submitted in October 2020.
[7] J. Lu, R. Goddard, K. Han and S. Hahn, 'Contact resistance between two REBCO tapes under load and load cycles', Supercond. Sci. Technol. **30** (2017) 045005.
[8] J. Qin *et al.*, 'New design of cable-in-conduit conductor for application in future fusion reactors', Superconduct. Sci. and Tech. **30** (2017) 115012.
[9] J.E. Menard *et al*, 'Fusion nuclear science facilities and pilot plants based on the spherical tokamak', Fusion **26** (2016) 106023